\documentclass[aps,pra,reprint, longbibliography,noeprint]{revtex4-2}

\usepackage{lipsum}

\usepackage{siunitx}
\DeclareSIUnit\gauss{G}
\usepackage{amsmath,bm}
\usepackage{dsfont}
\usepackage{amsbsy}
\usepackage{amssymb}
\usepackage{mathptmx, textcomp}
\usepackage{color}
\usepackage{braket}
\usepackage{graphicx}
\usepackage{textcomp}
\usepackage{notes2bib}
\usepackage{textcomp}
\usepackage{mwe}
\usepackage{siunitx}
\usepackage{physics}
\usepackage{filecontents}
\usepackage{soul}
\definecolor{bl}{rgb}{0, .1, .6}
\usepackage[colorlinks=true, citecolor = bl, linkcolor = bl, urlcolor=bl, pdfborder={0 0 0}]{hyperref}

\graphicspath{{./figures/}}

\usepackage[letterpaper,top=2cm,bottom=2cm,left=3cm,right=3cm,marginparwidth=1.75cm]{geometry}

\usepackage{amsmath}
\usepackage{graphicx}

\definecolor{giu}{rgb}{1,0.7,0.7}

\newcommand{\rb}[1]{{\color{red} #1}}

\begin{document}
\title{Narrowline cooling of dysprosium atoms in an optical tweezer array}
\author{Giulio Biagioni}
\author{Britton Hofer}
\author{Nathan Bonvalet}
\author{Damien Bloch}
\author{Antoine Browaeys}
\author{Igor Ferrier-Barbut}
\email{igor.ferrier-barbut@institutoptique.fr}
\affiliation{Universit\'e Paris-Saclay, Institut d'Optique Graduate School, CNRS, 
Laboratoire Charles Fabry, 91127, Palaiseau, France}

\begin{abstract}
    We perform narrowline cooling of single dysprosium atoms trapped in a 1D optical tweezers array, employing the narrow single-photon transition at 741 nm. At the trapping wavelength of 532 nm, the excited state is less trapped than the ground state. To obtain efficient cooling performances, we chirp the frequency of the cooling beam to subsequently address the red sidebands of different motional states. We demonstrate the effectiveness of the cooling protocol through Raman thermometry, which we characterize for our experimental conditions. We obtain an array of 75 atoms close to the motional ground state in the radial direction of the tweezers. Our results demonstrate the possibility to manipulate the motional degree of freedom of dysprosium in optical tweezers arrays, a key ingredient to exploit the potential of lanthanide-based tweezers platforms for quantum science.
\end{abstract}

\maketitle

\section{Introduction}
In single atom (molecule) platforms, such as optical tweezers and optical lattices, reaching a large motional ground state occupation is crucial for numerous tasks in quantum computing \cite{Bl22, Gr22, Hol23, Ba23, Blu24, Ru25} and quantum simulation \cite{Br20, Yo22, Ch22, Ch23}. Sideband-resolved cooling protocols are well-established for ions \cite{Mo95} as well as for alkaline and alkaline-earth(-like) atoms in optical tweezers \cite{Ka12, Th13, No18, Co18, Sa19} and in optical lattices \cite{Pe98, Vu98, Ha98}, and have been applied more recently to molecules \cite{Lu24, Ba24}. They have not yet been demonstrated in lanthanide species trapped in optical tweezers \cite{Bl23,Gr24} where one has to face strong anisotropic light shifts \cite{Lepers2014,Kao17,Becher2018,Chalopin2018a,Kreyer2021, Bl24, Pr24}, making the application of cooling techniques potentially more challenging. Complications may arise especially in tweezers arrays, in which polarization and intensity fluctuations directly affect the trapping homogeneity.
\\ 
Sideband-resolved cooling consists in addressing the red sideband, i.e.~the transition $n \rightarrow n-1$, to reduce the mean motional occupation number $\langle n \rangle$. The sideband-resolved regime can be reached either with a Raman transition involving two different states of the same manifold \cite{Ka12, Th13}, in which case an optical pumping beam is also required, or with a narrow single-photon transition satisfying $\Gamma < \omega$ \cite{No18, Co18, Sa19}, with $\Gamma$ the linewidth of the transition and $\omega$ the trapping frequency. In the second case, the cooling usually relies on magic trapping conditions, i.e.~the atoms are trapped with far-detuned light at carefully selected wavelengths, in which the differential light shift between the ground $\ket{g}$ and excited $\ket{e}$ states involved in the cooling transition vanishes. The magic condition can be complicated to realize experimentally, as it poses strict requirements regarding the trap wavelength. Moreover, the magic condition for the cooling light is likely to be incompatible with the magic condition for other transitions of interest, such as optical clock transitions \cite{Co19, Ur22}. Even if cooling in a non-magic trap is usually the result of experimental constraints, it involves interesting and qualitatively different cooling processes. Provided the excited state is trapped, an important consideration is whether the $\ket{e}$ state experiences a stronger confinement than the $\ket{g}$ state ($\omega_e > \omega_g$), or vice versa ($\omega_g > \omega_e$). In the first case, a fixed-frequency cooling scheme is still effective \cite{Co19, Ur22}, and the cooling mechanism was interpreted as an attractive Sisyphus cooling, in which the atom loses kinetic energy climbing the steeper trap of the $\ket{e}$ state, resulting in a favorable mismatch between the absorbed and spontaneously emitted photons of the cooling cycle \cite{Ta94, Iv11}. Also in this case a narrow linewidth is required, because the atom must first be excited preferentially at the bottom of the trap, and second it must have enough time to move in the excited state potential before decaying back. In the second case, i.e. when the excited state is less trapped than the ground state, a fixed-frequency cooling beam has been employed only to mitigate the recoil heating while imaging \cite{Co18}, but cannot produce a significant ground state population without substantial losses (see \cite{Ph24} for a recent theoretical comparison between different single-atom cooling mechanisms). Recently, a cooling scheme in the regime $\omega_e < \omega_g$ has been implemented with strontium \cite{Ho23}. It consists in chirping the laser frequency $\delta$ to continuously address the red sideband while the atom is cooled down, avoiding unwanted heating.\\ 

Here, we extend the narrow-line cooling toolbox to lanthanides, exploring the case of single dysprosium atoms in optical tweezers arrays. We perform cooling on the narrow transition at $\lambda = 741 $ nm, with linewidth $\Gamma/(2\pi) = 1.8$ kHz, between the manifolds $G = 4f^{10}6s^2$ $^5I_8$ and $E = 4f^{9}(^6H^o)5d6s^2$ $^5K^o_9$ \cite{Lu11}. The ground state trapping frequency in the tweezers radial direction during cooling is $\omega_g = 2\pi \times 28$ kHz. The narrow transition we use ensures the sidebands are resolved ($\omega/\Gamma>10$), despite the relatively low trapping frequency. The experiment is performed in tweezers light at 532 nm, with a (small) elliptical polarization fine-tuned to match the magic condition for single-atom imaging performed on the intercombination transition at 626 nm \cite{Bl24}. For the 741 nm transition, instead, we find that the tweezers light is not magic and the excited state is less trapped than the ground, $\omega_e < \omega_g$, realizing similar conditions as in \cite{Ho23}. We measure the final mean radial occupation number both with a release and recapture method and with Raman sideband resolved thermometry. After single-photon cooling, we find a decrease of factor 10 in the mean radial occupation number, reaching a 75 \% radial ground state occupation averaged over a 1D array of 75 traps, without losses nor post selection.

\section{Chirp cooling}\label{sec:2}

\begin{figure}
    \centering
    \includegraphics[width=\linewidth]{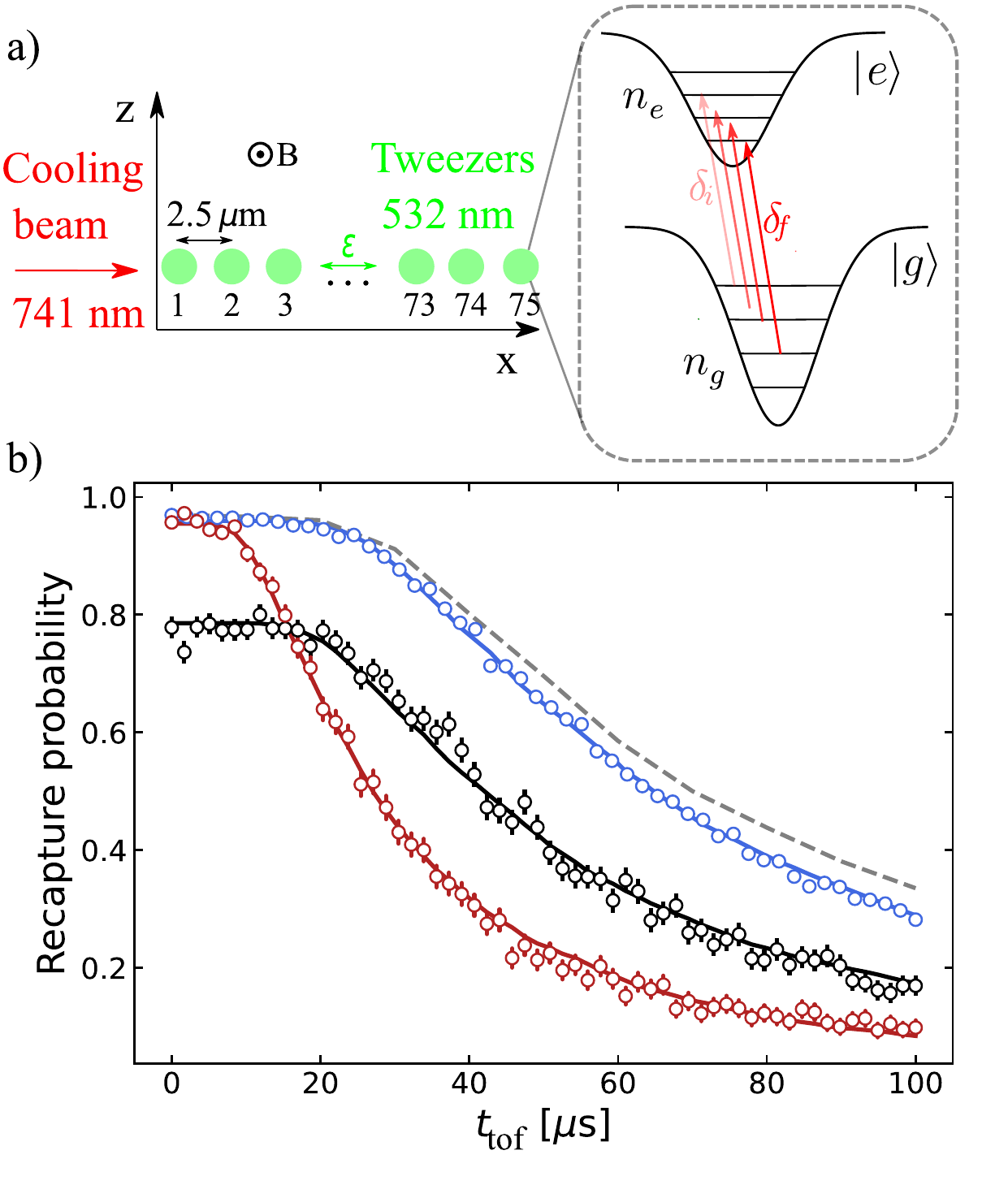}
    \caption{Chirp cooling and time of flight measurements. a) Sketch of the cooling protocol. We load a 1D array of 75 traps at 532 nm (green) propagating along $y$, with the main polarization axis $\epsilon$ along $x$. The cooling beam (741 nm) propagates along $x$ and the magnetic field is along $y$. We chirp the laser detuning from $\delta_i$ to $\delta_f$ to cool down the hottest atoms first and the coldest atoms last (inset). b) Release and recapture measurements. Survived fraction versus time of flight $t_{\text{tof}}$ without cooling (red), with a fixed-frequency cooling (black) and with chirp cooling (blue). Solid lines are classical simulations to extract the temperature, see text. The dashed line is a numerical simulation of the ground state release and recapture with the Schr\"{o}dinger equation, rescaled to 97.5 \% to take into account the imaging losses \cite{Bl23}.}
    \label{fig:1}
\end{figure}

The apparatus for trapping and imaging single dysprosium atoms is described in \cite{Bl23, Bl24, Ho24}. The cooling sequence starts with a 1D array of 75 traps spaced by $\SI{2.5}{\micro m}$ randomly loaded from a 3D Magneto Optical Trap (MOT) on the narrow intercombination transition at 626 nm. During the loading sequence, the tweezers have a trap depth of $\SI{120}{\micro K}$, while the atomic temperature after the loading is about $\SI{6}{\micro K}$. We lift the degeneracy between the Zeeman states with a magnetic field $\textbf{B} \sim (0,7,0)$ G aligned with the tweezers' propagation direction ($y$ axis, see Fig. \ref{fig:1}). The main direction of the tweezers polarization ellipse is along $x$. For the cooling protocol, we employ a trap depth corresponding to frequencies $(\omega_x, \omega_y, \omega_z)|_g = 2\pi \times (28, 4, 28)$ kHz. The two states involved in the cooling transition are $\ket{g} = \ket{G, J=8, m_J =-8}$ and $\ket{e} = \ket{E, J=9,m_J=-9}$. Although the ground state polarizability $\alpha_g$ at 532 nm has been measured \cite{Bl24}, the excited state polarizability $\alpha_e$ is experimentally unknown. We measure $\alpha_e/\alpha_g$ performing spectroscopy at different trap powers, see Appendix \ref{app:Polarizability}. In our experimental conditions, we find a ratio $\alpha_e/\alpha_g = 0.52 \pm 0.09$, and $\omega_e/\omega_g = \sqrt{\alpha_e/\alpha_g} = 0.72 \pm 0.06$ for the trap frequencies. We leave for future work the measurement of the scalar, vector and tensor components of the polarizability. However, we checked that the polarizability ratio only decreases by rotating the magnetic field or varying the ellipticity, so that the configuration with $\textbf{B}$ along the $y$ axis is the closest to magic conditions one can reach.\\

\textit{Fixed-frequency versus chirp cooling -} Cooling with a fixed frequency when $\alpha_e/\alpha_g <1$ causes losses due to unavoidable heating of a fraction of the atoms \cite{Co19}. To understand this, let us consider for simplicity only transitions involving the exchange of one motional quantum ($n_g \rightarrow n_e = n_g \pm 1$). Contrary to the magic trapping, now the resonance condition to address the red sideband, $n_e = n_g-1$, depends on the initial motional occupation number, $\delta = \omega_e(n_g-1)-n_g\omega_g$, where $\delta$ is the detuning of the laser calculated from the transition $n_g = 0 \rightarrow n_e = 0$, and from now on we indicate with $\omega_{g(e)}$ the trapping frequency in the radial direction for the ground (excited) state. For a given ratio $\omega_e/\omega_g = r < 1$, setting $\delta$ on the red sideband for some $n_g$ will, at the same time, make the beam resonant with the blue sideband ($n_e = n_g+1$) for a higher motional state $n_g' =n_g+2\,r/(1-r)$. The result is the cooling down of the coldest atoms and the heating up and loss of the hottest atoms. This mechanism was interpreted as a repulsive Sysiphus cooling: the cooling beam sets a repulsive threshold in energy space, above which atoms are pushed towards higher energy states, and vice versa for atoms below the threshold \cite{Co18}. \\

As theoretically proposed in \cite{Be21} and realized with strontium in \cite{Ho23}, chirping the frequency of the cooling laser allows one to cool down without losses. The chirp should first address $\Delta n>1$ transitions for the hottest atoms \cite{Yu18, Wu23} and then move closer to resonance to progressively address the red sidebands of the lower-$n_g$ states. \\

\textit{Release and recapture measurements -} In our setup, we employ a single cooling beam, tuned to the $\sigma^-$ transition between $\ket{g}$ and $\ket{e}$. We linearly ramp the detuning from $\delta_i$ to $\delta_f$ in a time $t_{\text{ramp}}$, at a constant intensity $I_c$. We then wait for $t_{\text{wait}}$ at the final detuning $\delta_f$, with the same intensity. To measure the final temperature, we perform a release and recapture experiment \cite{Tu08}. After the cooling stage, we switch the tweezers off and let the atoms evolve in free space for a variable time of flight $t_{\text{tof}}$, before switching the tweezers back on. We then take an image of the array and we extract the recapture probability. The experimental results are plotted in Fig. \ref{fig:1}(b). We compare three cases: chirp cooling, cooling at a fixed frequency and no cooling. We optimize the parameters $\delta_i, \delta_f, I_c, t_{\text{ramp}}$ and $t_{\text{wait}}$ (or just $\delta$ and $I_c$ for the fixed-frequency case) to reach the lowest temperature. Both the chirp and the fixed-frequency cooling are effective in lowering the temperature. When cooling with a fixed frequency, however, we lose 20\% of the atoms during the cooling stage, as manifested in the low recapture probability for $t_{\text{tof}} = 0$ $\si{\micro\second}$. We qualitatively explain the losses as resulting from the heating of the atoms sitting on the blue sidebands of the fixed-frequency beam, as explained above. Heating happens when we address $n_g\rightarrow n_e$ transitions with $n_e>n_g$. With the resonance condition for a $n_g\to n_e$ transition $\delta = n_e\,\omega_e-n_g\,\omega_g$ and our experimental detuning $\delta=-1.6\,\omega_g$, we expect to increase the motional quantum number ($n_e>n_g$) for atoms with $n_g^{h} = 5^{+1}_{-0}$. Assuming an initial thermal state, the heated fraction is then $\sum_{n>n_g^h} P_n \approx 25 \%$, where $P_n = (1-\exp(-\hbar\omega_g/k_BT))\exp(-n\hbar\omega_g/k_BT) $ is the probability that the atom occupies the level $n$, in relatively good agreement with the experiment. \\ 
When applying chirp cooling, instead, we do not observe such losses. The data presented in Fig. \ref{fig:1}(b) are with $\delta_{i}/\omega_g = -11$, $\delta_{f}/\omega_g = -1.1$, $t_{\text{ramp}} = 70$ ms, $t_{\text{wait}} = 5$ ms and $I_c /I_s = 62$, with $I_s = 0.6$ $\si{\micro\watt}/\si{\centi\metre}^2$ the saturation intensity. When bringing the final detuning $\delta_f$ closer to resonance, we observe losses since the cooling light becomes resonant with the carrier transition from the ground state. We also found similar cooling performances with faster ramps, up to $t_{\text{ramp}} = 30 $ $\si{\milli\second}$, employing the same ramp speed and an initial detuning closer to resonance. Instead, even faster ramps lead to larger final temperatures.
\\ 

\textit{Classical and quantum simulations of the release and recapture - }To extract the temperature from the experimental data, we perform classical simulations of the dynamics \cite{Tu08}. We sample initial positions $\textbf{r}_i$ and velocities $\textbf{v}_i$ from a Boltzmann distribution at temperature $T$ and with the external potential $V_{\text{trap}}$ given by the gaussian profile of the tweezers. The latter is independently calibrated \cite{Bl24}. We then evolve the positions $\textbf{r}_i (t) = \textbf{v}_i t + \textbf{r}_i (0)$ in time and we consider that the atom is recaptured if its kinetic energy at $t=t_{\text{tof}}$ is smaller than the potential energy $V_{\text{trap}}(\textbf{r}_i (t_{\text{tof}}))$, characterized by the tweezers' waist and depth. We average over about 5000 repetitions. We then fit the experimental data with the simulation result, taking the temperature $T$ and the initial recapture probability as the only free parameters. We get $T = 6.3(1)$ $\si{\micro\kelvin}$ for the non-cooled sample, $T = 2.4(1)$ $\si{\micro\kelvin}$ with the fixed frequency cooling and $T= 1.6 (1)$ $\si{\micro\kelvin}$ with the chirp cooling. From the fitted data we can extract the radial mean occupation number $\langle n_r \rangle$ through

\begin{equation}\label{eq:mon}
    \langle n_r \rangle = \frac{1}{e^{\hbar\omega_g/k_B T}-1}.
\end{equation}

We perform the release and recapture measurement by increasing the trap power after cooling, up to a value corresponding to $\omega_g = 2\pi \cross 45$ kHz (different from the cooling one), which is then the relevant parameter in Eq. \ref{eq:mon}. We get a decrease in radial occupation number from $\langle n_r \rangle = 2.45(10)$ before cooling to $\langle n_r \rangle = 0.36(5)$ after cooling. \\
For comparison with the data, we also compute the recapture probability solving the time-dependent Schr\"{o}dinger equation with the 3D ground state as the initial state as in \cite{Ho23}. We expand in free flight the wave function for $t_{\text{tof}}$ and then we switch on the 3D tweezers potential. After a fixed additional evolution time (100 $\si{\micro\second}$), we compute the recapture probability as the fraction of the wavefunction inside the tweezer potential. The result is plotted in Fig. \ref{fig:1}(\st{c}\rb{b}), showing that the experimental data is indeed very close to the ground state.\\

\begin{figure}
    \centering
    \includegraphics[width=\linewidth]{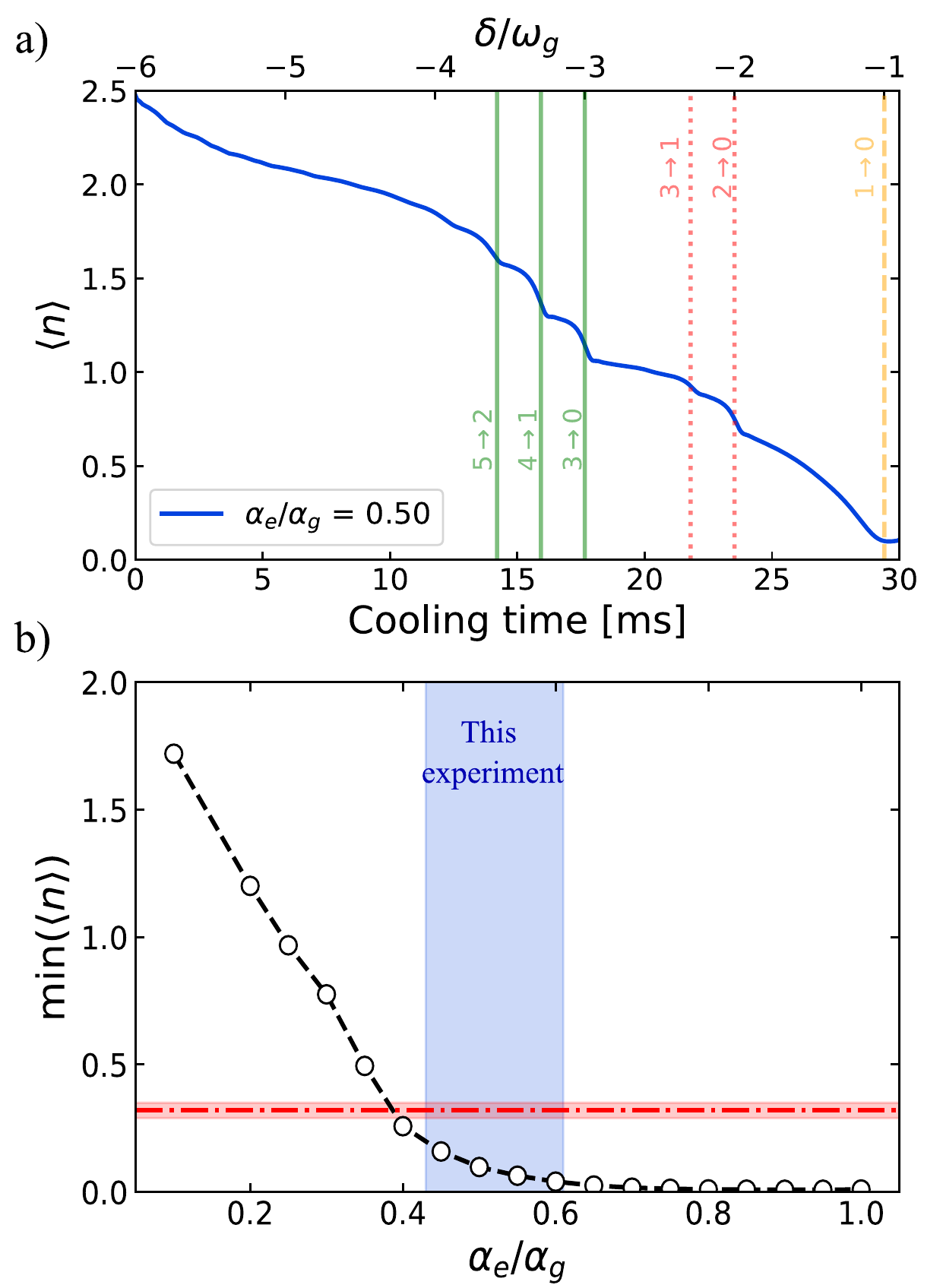}
    \caption{Simulations of the chirp cooling. a) Mean occupation number as a function of cooling time during a frequency ramp from $\delta_i/\omega = -6$ to $\delta_f/\omega = -0.9$ (upper horizontal axis) in $t_{\text{ramp}} =$ 30 ms, for $\alpha_e/\alpha_g = 0.5$. The initial state is thermal with $\langle n \rangle=2.5$. The vertical lines indicate motional transitions corresponding to faster-decreasing steps in $\langle n \rangle$. Thick, dotted and dashed lines indicate $n \rightarrow n-3$, $n \rightarrow n-2$, and $n \rightarrow n-1$ transitions, respectively. Despite being reduced when the Lamb-Dicke parameter $\eta$ is small, transitions with $\Delta n>1$ are still significant when $\langle n\rangle$ is large \cite{Yu18, Wu23, Ho23}. (b) Minimum occupation number reached for different ratios $\alpha_e/\alpha_g$. The initial and final detunings of the ramp are the same as in a), and $t_{\text{ramp}}$ is either 30 ms or 50 ms to let the system reach the minimum occupation number. The blue region indicates the uncertainty interval for the polarizability ratio in our experiment. The red region indicates the Raman thermometry measurement averaged over the whole array.}\label{fig:2}
\end{figure}

\textit{Simulations of the chirp cooling - } We model the chirp cooling ramp using the 1D hamiltonian in \cite{Ho23}

\begin{equation}\label{hamiltonian}
    \hat{H} = \hat{H}_{\text{ho}} + \hat{H}_{\text{drive}}.
\end{equation}

The first term is the harmonic oscillator hamiltonian $\hat{H}_{\text{ho}} = \hbar\omega_g(\hat{a}^\dagger\hat{a}+1/2) + \frac{\hbar(\omega_e^2-\omega_g^2)}{4\omega_g}(\hat{a}+\hat{a}^\dagger)^2\ket{e}\bra{e} + U\ket{e}\bra{e}$, taking into account the state-dependent trap frequencies. The energy $U = \hbar(\omega_g-\omega_e)/2$ is chosen such that the transition $n_g =0 \rightarrow n_e = 0$ is driven resonantly with $\delta = 0$. The second term is the drive hamiltonian in the rotating wave approximation $\hat{H}_{\text{drive}} = \hbar(\Omega e^{i\eta (\hat{a}^\dagger+\hat{a})}\ket{e}\bra{g} +h.c.)/2 - \hbar\delta(t)\ket{e}\bra{e}$, with $\Omega$ the Rabi frequency and $\eta = \frac{2\pi}{\lambda}\sqrt{\frac{\hbar}{2m\omega_g}} \approx 0.3$ the Lamb-Dicke parameter, with $m$ the $^{162}$Dy mass. The main difference from sideband cooling in magic traps is the additional term in the harmonic hamiltonian, quadratic in the creation and annihilation operators $\hat{a}^\dagger$ and $\hat{a}$. This term induces $n\rightarrow n\pm 2$ transitions regardless of the Lamb-Dicke parameter \cite{Ph24}. We include the spontaneous emission from the excited state with rate $\Gamma$ through the jump operators $\hat{L}_{\theta} = \sqrt{\frac{\hbar\Gamma}{2}}e^{i\eta(\hat{a}^\dagger+\hat{a})\cos{\theta}}\ket{g}\bra{e}$ describing the emission of a photon at an angle $\theta \in [0,\pi]$ with respect to the trapping direction. We then solve the master equation for the density matrix $\rho$, $\dot{\rho} = -\frac{i}{\hbar}[\hat{H},\rho]+\int \big(\hat{L}_\theta\rho\hat{L}_\theta^\dagger-\frac{1}{2}\{\hat{L}_\theta^\dagger\hat{L}_\theta,\rho\})\sin{\theta}\text{d}\theta$ with QuTip \cite{Jo13}. We employ 20 harmonic oscillator levels. In Fig. \ref{fig:2} we plot the mean occupation number $\langle n \rangle$ during a cooling ramp from $\delta_i/\omega_g = -6$ to $\delta_f/\omega_g = -0.9$ in $t_{\text{ramp}} = 30$ ms, with $\alpha_e/\alpha_g = 0.5$. Starting from a thermal state with $\langle n \rangle = 2.5$, we reach $\langle n \rangle = 0.18$ at the end of the ramp. We repeat the same simulations for different polarizabilty ratios, and we plot the minimum of $\langle n \rangle$ at the end of the ramp in Fig. \ref{fig:2}(b). Our measurements averaged over the 75 traps (red line) give a sligthly larger $\langle n \rangle$ than the 1D single-particle simulations. This is due to site-by-site intensity variations across the tweezers array that make the cooling efficiency trap-dependent, as we discuss in Section \ref{sec:4}.

\section{Thermometry with Raman spectroscopy}\label{sec:3}

The mean occupation number extracted from the release and recapture method is expected to overestimate the actual one when approaching the ground state. The classical simulations, indeed, neglect the zero-point energy, which appears as a fictitious minimum detectable temperature of $T_{gs} = \hbar\omega_g/2k_B = 1.1$ $\si{\micro\kelvin}$ in the release and recapture sequence. Moreover, the analysis is sensitive to the trap parameters (waist and intensity), which can be hard to measure precisely. We can get a more reliable measurement of the mean occupation number from the ratio between the red and blue sidebands of the spectrum \cite{Th13, Ka12}. Since the transition is not magic, however, intensity fluctuations across the trap array (see next Section) prevent us from resolving the sidebands in the averaged spectrum. Therefore, we rely on Raman sideband resolved spectroscopy to perform thermometry of the cooled array.\\

\begin{figure}
    \centering
    \includegraphics[width=\linewidth]{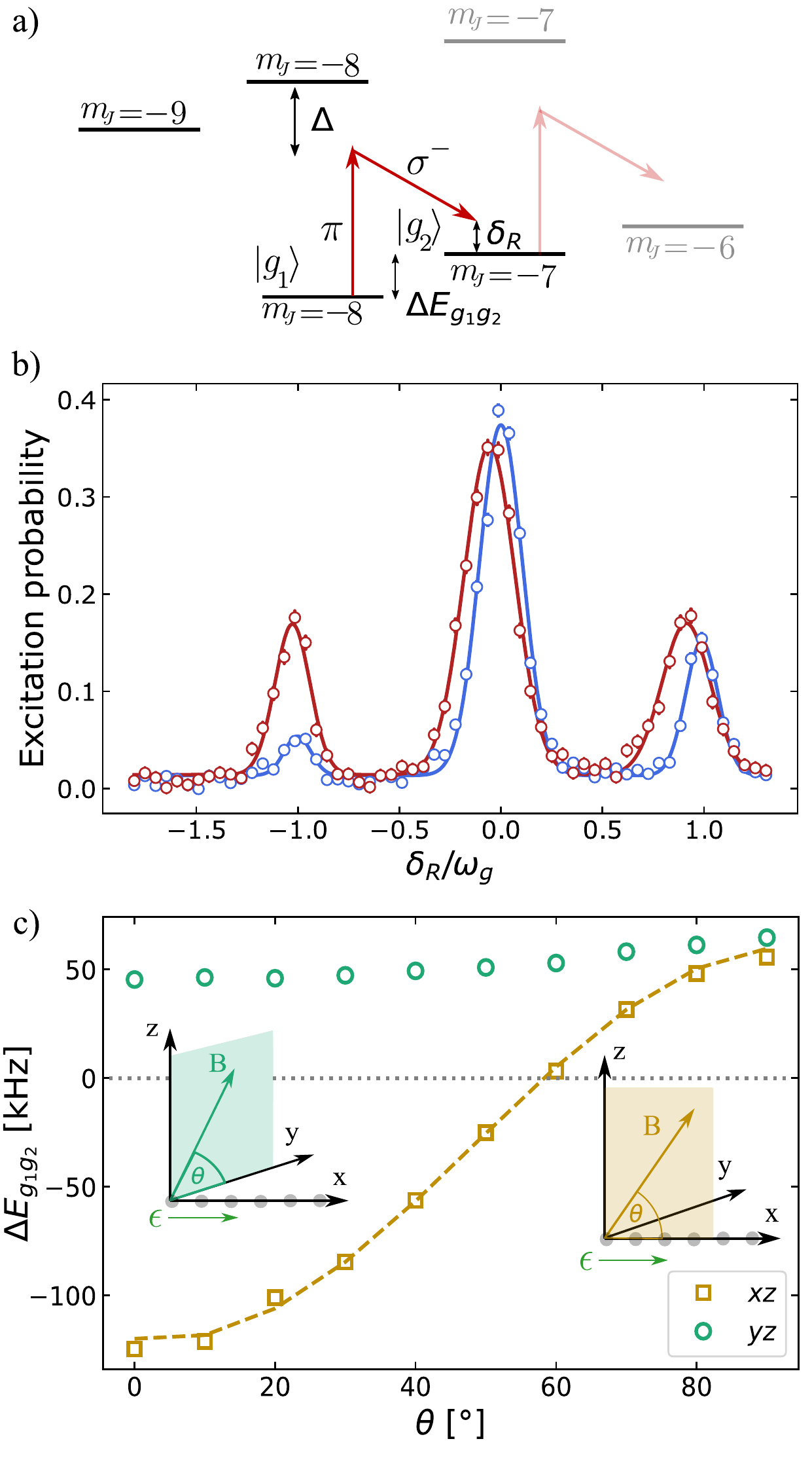}
    \caption{Raman thermometry. a) Sketch of the Raman transition and energy levels involved, see text. Energies are not in scale for clarity (in the experiment, $\Delta \gg \delta_R$), and the Zeeman splitting is not depicted to highlight the light shift energy $\Delta E_{g_1g_2}$. b) Raman sideband resolved spectra without cooling (red) and after the chirp cooling (blue), together with gaussian fits (solid lines) to extract the sideband populations. c) Light shift of the carrier transition as a function of the magnetic field direction in the $yz$ (green circles) and $xz$ plane (yellow squares), with $P_T = 1.8$ mW. The yellow dashed line is a $\cos^2{\theta}$ fit, while the gray dotted line indicates the magic angle $\theta_m$.}
    \label{fig:3}
\end{figure}

\textit{Raman sequence -} We drive a two-photon Raman transition between two Zeeman states of the ground manifold, $\ket{g_1} = \ket{G, J = 8, m_J =-8}$ and $\ket{g_2} = \ket{G, J = 8, m_J =-7}$, see Fig. \ref{fig:3}. The two beams propagate in the radial plane of the tweezers, hence the momentum exchanged with the atoms $\Delta \bf{k} = \bf{k_1}-\bf{k_2}$ is also in the radial plane. After the cooling stage, we address the Raman transition with a detuning from the excited state of $\Delta \sim$ 4 MHz $= 2\times 10^3 \Gamma$, and we scan the detuning $\delta_R$ from the transition $\ket{g_1} \rightarrow \ket{g_2}$, see Fig. \ref{fig:3}(a). The two-photon Rabi frequency is typically in the range 1-10 kHz, low enough to resolve the radial sidebands. To detect the atoms transferred in $\ket{g_2}$, we apply a long pulse with $\sigma^-$ polarization resonant with the $\ket{g_1} \rightarrow \ket{E, J=9, m_J =-9}$ transition, but detuned from the $\ket{g_2} \rightarrow \ket{E, J=9, m_J =-8}$ transition by 2 MHz thanks to the difference in Lande factors between the $G$ and $E$ manifolds (1.241 an 1.22 respectively, \cite{Lu11}). The pulse causes losses of $\ket{g_1}$ atoms due to heating, but keeps $\ket{g_2}$ atoms unaffected, given the narrow linewidth of the transition. Finally, we optically pump $\ket{g_2}$ atoms back into $\ket{g_1}$ with a pulse resonant on the $\sigma^-$ transition at 626 nm, and we image them. Each of these steps is independently optimized by tuning the tweezers' power and the applied magnetic field. In particular, we perform the Raman transition with a tweezers power $P_T = 1.8$ mW, corresponding to trap frequencies $(\omega_x, \omega_y, \omega_z)|_g = 2\pi \times (45, 7, 45)$ kHz.\\

In Fig. \ref{fig:3}(b) we compare the Raman spectrum with and without the chirped cooling protocol. We clearly resolve the sideband associated with radial motional transitions, and we observe a reduction of the red sideband when performing the chirp cooling. Fitting the probability to excite the red and blue sidebands, $P_R$ and $P_B$ \footnote{We identify $P_R$ and $P_B$ as the area of the red and blue sideband peaks, respectively.}, we can compute the average radial occupation number without free parameters
\begin{equation}
    \langle n_r \rangle = \frac{P_R/P_B}{1-P_R/P_B}.
\end{equation}
From the measurement in Fig. \ref{fig:3}, we get $\langle n_r \rangle = 2.5 \pm 0.4$ before cooling and $\langle n_r \rangle = 0.32 \pm 0.03$ after cooling, in agreement with the release and recapture experiments. The corresponding ground state population is $P_0 =1/(1+\langle n_r \rangle) = 76(2) \%$.\\ 
It is natural to wonder whether it is possible to further cool down through Raman sideband cooling and optical pumping back from $\ket{g_2}$ to $\ket{g_1}$. We tried to perform the optical pumping step both with the 626 nm (magic) and the 741 nm (non-magic) transition, but without observing a further reduction in temperature. This is probably due to the inconvenient ratio of the Clebsh-Gordan coefficients for the two decay channels  $m_{J'}=-8\rightarrow m_{J}=-7$ and $m_{J'}=-8\rightarrow m_{J}=-8$ involved in the optical pumping process, equal to $\abs{C_{-8\rightarrow-8}/C_{-8\rightarrow-7}}^2 = 0.12$. As a consequence, one atom needs to scatter on average $\sim$8 photons before being pumped back in $\ket{g_1}$. In each scattering event, the probability of heating the atom through the transition $n\rightarrow n+1$ is $\eta^2 (n+1) \sim 0.05(n+1)$, so on average this heating overcomes the Raman cooling \footnote{Here, the Lamb-Dicke parameter $\eta\approx 0.22$ is lower than in the chirp cooling.}. We leave the exploration of alternative optical pumping mechanisms for future work.\\
In the axial direction, we expect the cooling protocol to be less effective, due to the larger Lamb-Dicke parameter $\eta_{\text{axial}} = 0.75$. We performed Raman thermometry in the axial direction, rotating one of the Raman beams so that $\Delta k$ has a component along $y$. We observed sidebands up to the 5th order, but without a clear reduction of the red sidebands after cooling. In the future, increasing the axial confinement, for example with an optical lattice, will allow us to apply the cooling protocol also in the axial direction.\\

\textit{Light shift of the Raman transition -} The two states $\ket{g_1}$ and $\ket{g_2}$ involved in the Raman transition both belong to the $G$ manifold, so they have the same scalar polarizability $\alpha_g^s$. However, they still experience a different light shift due to the vector ($\alpha_g^v$) and tensor ($\alpha_g^t$) polarizabilities \cite{Bl24}, which produce an $m_J-$dependent light shift. The vector light shift is non-zero only if the light polarization is elliptical. In our case, the tweezer light has a small ellipticity angle of $6.5^\circ$, which, together with a small $\alpha_g^v$ (compatible with zero \cite{Bl24}), makes the vector light shift much smaller than the tensor one. The main features of the light shift on the Raman transition can then be understood by assuming a linear polarization along the $x$ axis. In this case, the light shift energy is $E_{LS}(m_J) = -\frac{I}{2\epsilon_0 c}\alpha(m_J)$, with the total polarizability
\begin{equation}\label{eq:polarizability}
   \alpha(m_J) =  \alpha_g^s+\alpha_g^t[3(\bm{b}\cdot\bm{x})^2-1]\Big(\frac{3m_J^2-J(J+1)}{2J(2J-1)}\Big),
\end{equation}
where $\bm{b}$ the magnetic field direction. We measure the light shift of the carrier transition $\Delta E_{g_1g_2} = E_{LS}(-8)-E_{LS}(-7)$ by changing the magnetic field angle $\theta$ with respect to the $xy$ plane, for a fixed magnitude $B=3.3$ G, see Fig. \ref{fig:3}(c). When $\textbf{B}$ is in the $xz$ plane, $\bm{b}\cdot\bm{x} = \cos{\theta}$. We then observe the typical $\sim(3\cos^2{\theta}-1)$ behavior of the tensor light shift \cite{Bl24}, with a magic angle at $\theta_m \approx 54.7^\circ$. When $B$ is instead in the $yz$ plane, $\bm{b}\cdot\bm{x} = \cos{\pi/2}=0$ so the tensor light shift is fixed. The residual weak oscillation can be due to a weak vector light shift or to errors in the magnetic field calibration, resulting in a small varying $B$ component in the $xz$ plane while $\theta$ is scanned. \\

The magnitude of the light shift determines the number of Zeeman levels involved in the Raman transition, see Fig. \ref{fig:3}(a). At the magic angle, we expect to populate also the Zeeman levels with $m_J > -7$ since the Raman transition is resonant with any $m_J \rightarrow m_J +1$ transition. On the other hand, when the light shift is large enough, the Raman beams only couple the two states $\ket{g_1}$ and $\ket{g_2}$. We took Raman spectra in both configurations, without observing a difference in the temperature obtained. In particular, the data in Fig. \ref{fig:3}(b) are for $B$ along the $z$ axis ($\theta = \pi/2$). \\

We however observe two effects of the light shift on the Raman spectra. The first is a trap-dependent light shift arising from a polarization gradient across the tweezers' array, which we investigate in Appendix \ref{app:Raman slopes}. The second is the redshift of the Raman spectrum at higher temperatures, visible in the data of Fig. \ref{fig:3}(b). It is around 3 \%, 6 \% and 8 \% of the trap frequency $\omega_g$ for the red sideband, carrier and blue sideband respectively. It arises from the different trap frequencies $\omega_{g_1}$ and $\omega_{g_2}$ that make the transition $\ket{g_1} \rightarrow \ket{g_2}$ slightly non-magic. As such, the position of the three peaks weakly depends on the initial occupation number, which is analogous to what we discussed for the $\ket{g} \rightarrow \ket{e}$ transition in the chirp cooling protocol. For a state $n$, the Raman peaks are at $\delta_R^n/\omega_{g_1} = (n +i)\sqrt{\alpha_{g_2}/\alpha_{g_1}}- n$, with $i = \{-1,0,1\}$ for the red sideband, carrier and blue sideband, respectively. Computing $\alpha_{g_1} = \alpha(-8)$ and $\alpha_{g_2} = \alpha(-7)$ from Eq. \ref{eq:polarizability}, with $J=8$ and $\theta = \pi/2$, we have $\sqrt{\alpha_{g_2}/\alpha_{g_1}} = \sqrt{(1-5\alpha_g^t/16\alpha_g^s)/(1-\alpha_g^t/2\alpha_g^s)} \approx 1+3\alpha_g^t/32\alpha_g^s$ at first order in $\alpha_g^t/\alpha_g^s$. Since the two components of the polarizability have opposite sign \cite{Bl24}, $\alpha_g^s>0$ and $\alpha_g^t<0$, the factor $\sqrt{\alpha_{g_2}/\alpha_{g_1}}<1$ and the Raman resonances $\delta_R^n$ are red-shifted proportionally to $n$. Using the available measurement of the ground state polarizabilities \cite{Bl24}, we get the correct order of magnitude $|\delta_R^n -\delta_R^0|/\omega_{g_1}\approx 3\%$ for our hot spectrum with $\langle n \rangle = 2.5$. A more quantitative comparison would require a more accurate knowledge of the polarizabilities, at the percent level. We also checked that the shifts vanish when addressing the transition at the magic angle $\theta_m$.

\section{Temperature variations along the array}\label{sec:4}

\begin{figure}
    \centering
    \includegraphics[width=\linewidth]{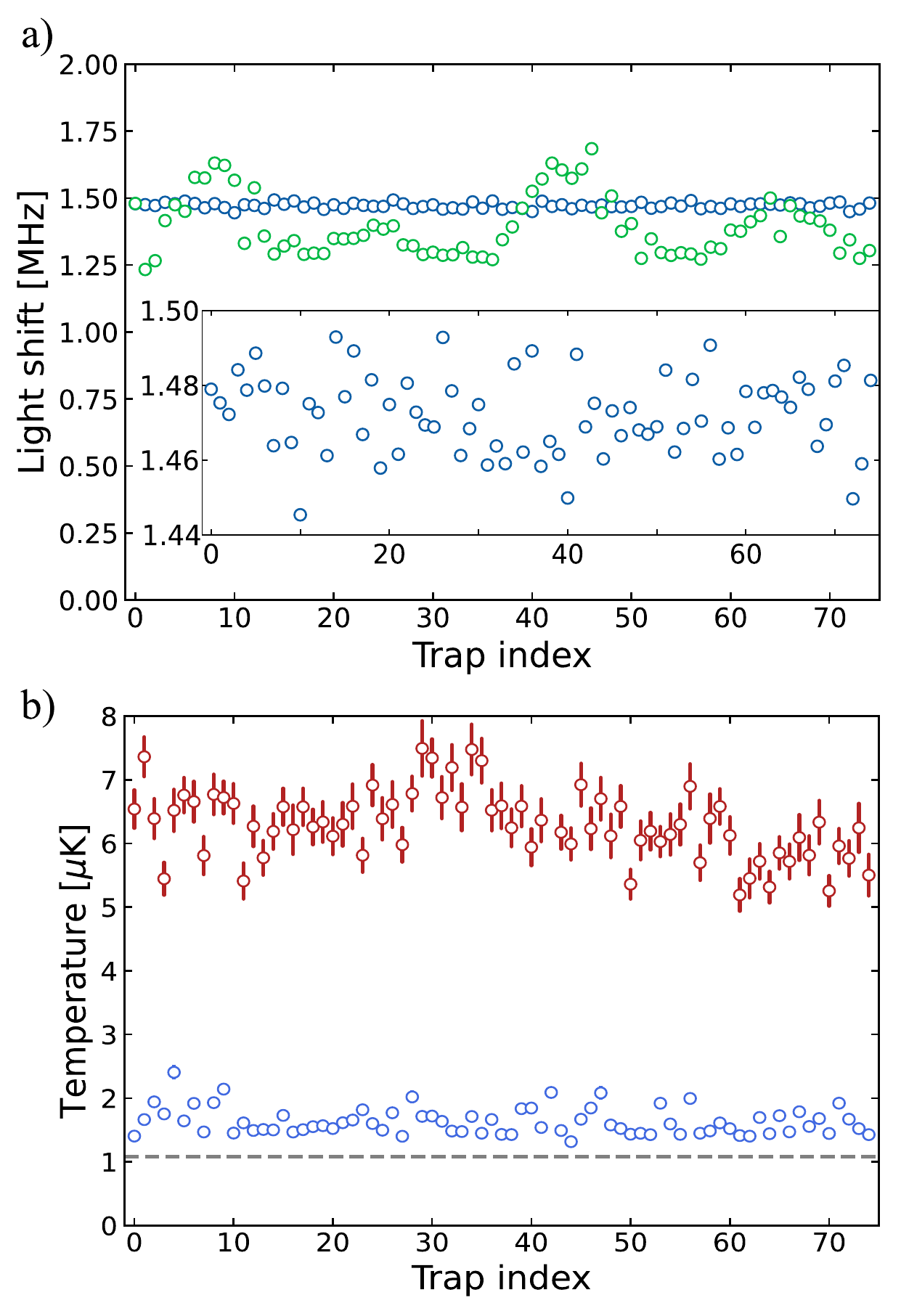}
    \caption{Homogeneity of the cooling across the array. (a) Light shift of the $\ket{g}\rightarrow \ket{e}$ as a function of the trap position, before (green) and after (blue) the homogenization algorithm, for $P_T = 2.2$ mW. The inset shows a zoom for the homogenized array. (b) Temperature $T$ from a release and recapture experiment as a function of the trap position. Red points are without cooling and blue points are after chirp cooling performed with $P_T = 0.7$ mW. The horizontal dashed line is the energy associated with the ground state, $T_{gs} = \hbar\omega_g/2k_B$, with $\omega_g = 2\pi\times$ 45 kHz the radial trap frequency in the release and recapture sequence.}
    \label{fig:4}
\end{figure}

All the datasets shown in Figs. \ref{fig:1}-\ref{fig:3} are averaged over the whole chain of 75 traps. When the traps are not magic, any variations in the tweezers' parameters (intensity, waist, polarization, etc.) from one trap to another correspond to variations in the energy separation between $\ket{g}$ and $\ket{e}$. To efficiently cool all the atoms in the 1D chain, we need these variations to be smaller than the radial trapping frequency. We therefore implemented a homogenization algorithm that acquires an image of the tweezers array on a service camera and modifies the radio-frequency tones of the acousto-optic deflector (AOD) to correct for variations in the optical intensity. We use the trap-resolved light shift as weight for the homogenization procedure. After a few iterations of the algorithm, we reduce the intensity variations over the trap array, defined as the standard deviation of the light shift, from 7 \% to 0.7 \% of its mean value, see Fig. \ref{fig:4}(a).\\

To check the degree of temperature homogenization over the array, we perform the same release and recapture measurements analysis as in Fig. \ref{fig:1}(b), but trap by trap. The fitted temperatures $T$ as a function of the trap index are reported in Fig. \ref{fig:4}(b), with and without having applied the cooling sequence. We obtain a cold sample with the variations across the 75 traps of the array having standard deviation $\sigma_{T} =\SI{0.2}{\micro\kelvin}$. We do not observe any correlation between final and initial temperatures, indicating that we correctly address the red sidebands also of the hotter atoms at the beginning of the cooling ramp.\\

The residual intensity variations after homogenization still limit the cooling performance over the array. At low tweezers power $P_T$, the light shift fluctuations are smaller but so is the trap frequency, rendering chirp cooling less efficient. On the other hand, at large $P_T$ it is easier to resolve the sidebands but the intensity variations are larger, and the traps reach different final temperatures. Since the light shift scales linearly with $P_T$ while the trap frequency scales as $\sqrt{P_T}$, there is an optimal value of $P_T$ for which the cooling is efficient for all traps. We find this optimal value to be $P_T = 0.7$ mW, corresponding to a radial trap frequency of $\omega_g = 2\pi \cross 28$ kHz as stated in Section \ref{sec:2}. The measurements in Fig. \ref{fig:4}(b) correspond to $P_T = 0.7$ mW.\\

\section{Conclusions}
We have demonstrated narrowline cooling for an array of single dysprosium atoms in the radial direction of non-magic optical tweezers. This cooling allowed for a ground-state population of 76(2) \%. We reached homogeneous cooling over 75 traps with standard deviation $\sigma_{T} = 0.2$ $\si{\micro\kelvin}$ which validates its use in many-atom ensembles. Our results extend narrow-line cooling protocols previously employed for alkaline-earth (like) atoms to lanthanides, where one benefits from several transitions from the ground state with different linewidth, suited for different laser cooling protocols.  Narrow-line cooling will find applications also in existing setups of single erbium atoms in optical tweezers \cite{Gr24} and optical lattices \cite{Su23}. 

\begin{acknowledgments}
We thank Yannis Bencherif and Yann Le Cor for assistance on the experiment and simulations.
This project has received funding by the Agence Nationale de la Recherche (ANR-22-PETQ-0004 France 2030, project QuBitAF), by the European Union (ERC StG CORSAIR,  101039361, ERC AdG ATARAXIA 101018511), and the Horizon Europe programme HORIZON-CL4- 2022-QUANTUM-02-SGA (project 101113690 PASQuanS2.1).
\end{acknowledgments}

\medskip
All data corresponding to the findings of this manuscript are available in \cite{dataset}.

\appendix

\section{Excited-state polarizability}\label{app:Polarizability}

Here we show our measurement of the ratio $\alpha_e/\alpha_g$, which was experimentally unknown for the 741 excited state. Theoretical calculations are notoriously difficult for lanthanides, due to the abundance of transitions that contribute to the polarizability in spectral regions not dominated by a single transition \cite{Dz11, Li16}. Moreover, due to their open-shell electronic configuration, lanthanides typically have large tensor and vector contributions to the polarizability, so the trapping potential can be very sensitive to the relative orientation between the magnetic field $\textbf{B}$ and the tweezers polarization, see Section \ref{sec:3} and Appendix \ref{app:Raman slopes}.\\

\begin{figure}[b]
    \centering
    \includegraphics[width=\linewidth]{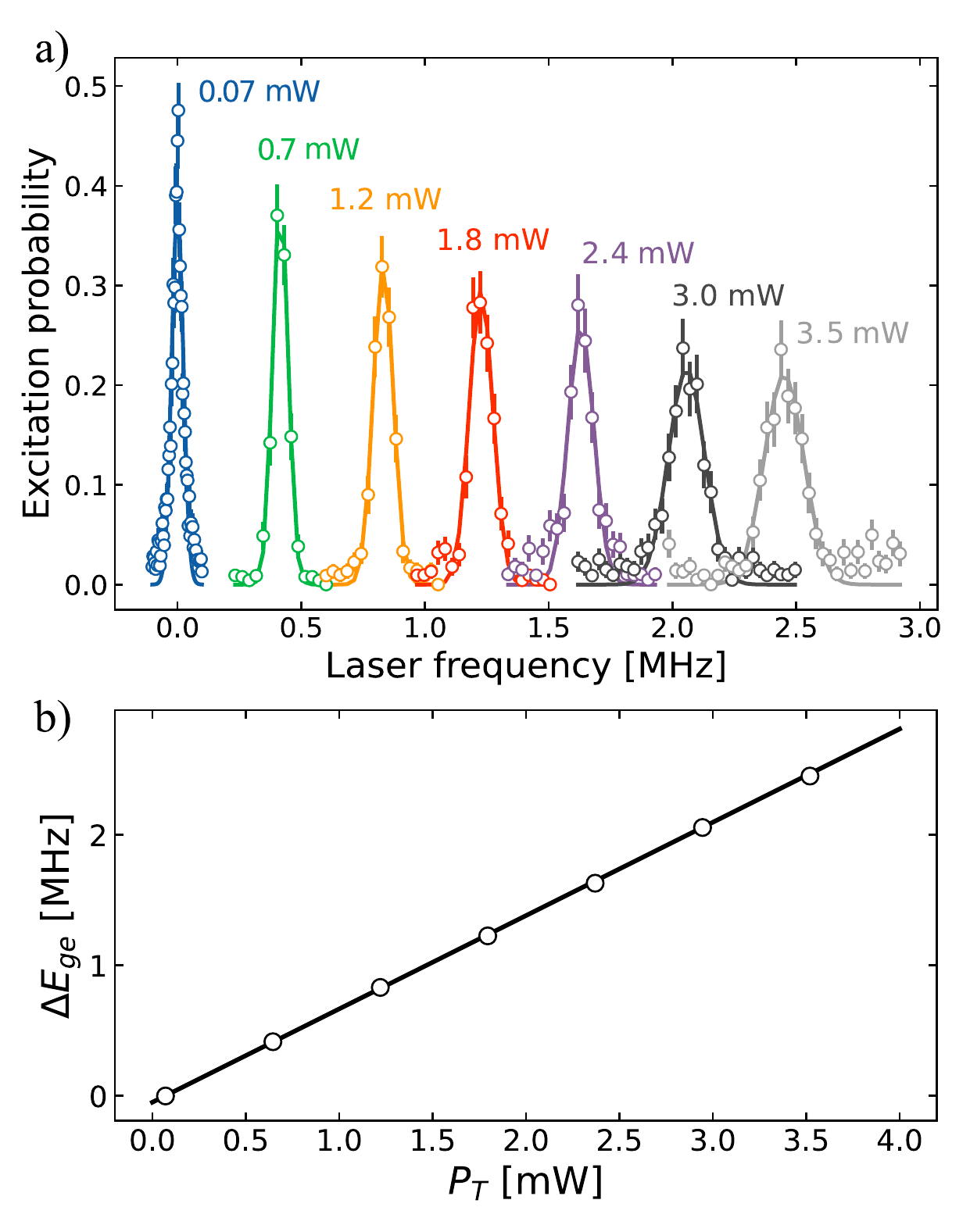}
    \caption{Measurement of the excited state polarizability. a) Spectroscopy of the 741 nm transition in different 532 nm trap depths, corresponding to different power per tweezer $P_T$. Thick lines are gaussian fit to the data to extract the corresponding light shifts $\Delta E_{ge}$. b) Light shift $\Delta E_{ge}$ fitted from the spectra in a) versus tweezer power $P_T$. The black line is a linear fit.}
    \label{fig:app_1}
\end{figure}

We estimate the ratio $\alpha_e/\alpha_g$ in the same conditions as used for the cooling protocol ($\bf{B}$ oriented in the $\bf{y}$ direction, corresponding to $\theta = \pi/2$), by performing spectroscopy on the 741 transition at different trap depths. We follow the spectroscopy protocol outlined in \cite{Ho24}. We shine a  pulse of $\sigma^-$ polarized light resonant with the cycling transition between $\ket{g} = \ket{G, J=8, m_J =-8}$ and $\ket{e} = \ket{E, J = 9, m_J =-9}$, reaching the steady state. Just after the end of the pulse, we switch on a fast (100 ns) beam resonant on the strong transition at 421 nm (linewidth $2\pi\cross$ 32 MHz). This beam removes the atoms in $\ket{g}$ while atoms in $\ket{e}$ are unaffected. Crucially, the state $\ket{e}$ doesn't decay significantly during this time (the lifetime of $\ket{e}$ is 88 $\si{\micro\second}$). Only after the pulse, excited atoms decay back to $\ket{g}$ and are imaged. The results of the spectroscopy are displayed in Fig. \ref{fig:app_1}. Since the resonance shifts with the tweezers power, the trapping potential is not magic for the 741 transition. The light shift energy of each state is given by $E_{LS} = -\alpha I/(2\epsilon_0 c)$, with $I$ the intensity of the tweezer and $\alpha$ is the total polarizability. The differential light shift between the ground and excited state is thus $\Delta E_{ge} = E_{LS}^e - E_{LS}^g = (\alpha_g -\alpha_e)I/(2\epsilon_0 c)$. Since the resonance gets blue-shifted while increasing the tweezers power, $\alpha_e < \alpha_g$. \\

To measure the ratio $\alpha_e/\alpha_g$, we fit the centers of the spectroscopic peaks as a function of the tweezers power $P_T$, obtaining the light shift $\Delta E_{ge} (P_T)$, displayed in Fig.\ref{fig:app_1}. We then write the light shift as $\Delta E_{ge} = (1-\alpha_e/\alpha_g)U_g$, with $U_g = \alpha_g I/(2\epsilon_0 c)$ the trap depth experienced in the ground state. In \cite{Bl24} we have measured $U_g = k_B \times (119\pm7)$ $\si{\micro\kelvin}$ for a trap power $P_T = (1.7 \pm 0.3)$ mW, with the same magnetic field orientation \st{$\theta = \pi/2$} as for the data in Fig. \ref{fig:app_1}. We obtain a consistent value for $U_g$ also from the fit of the tensor light shift in the Raman transition, Fig. \ref{fig:3}(c). From the linear fit of the light shift energy versus tweezers power in Fig. \ref{fig:app_1} we get $\Delta E_{ge}((1.7\pm0.3) \text{ mW}) = 2\pi \times (1.2 \pm 0.2)$ MHz, and solving for the polarizability ratio we get $\alpha_e/\alpha_g = 0.52 \pm 0.09$. We plot this uncertainty interval in Fig. \ref{fig:2} to compare the simulations of the chirp cooling with the measured temperature.\\

\section{Trap-dependent light shift in the Raman spectroscopy}\label{app:Raman slopes}

\begin{figure*}
    \centering
    \includegraphics[width=\linewidth]{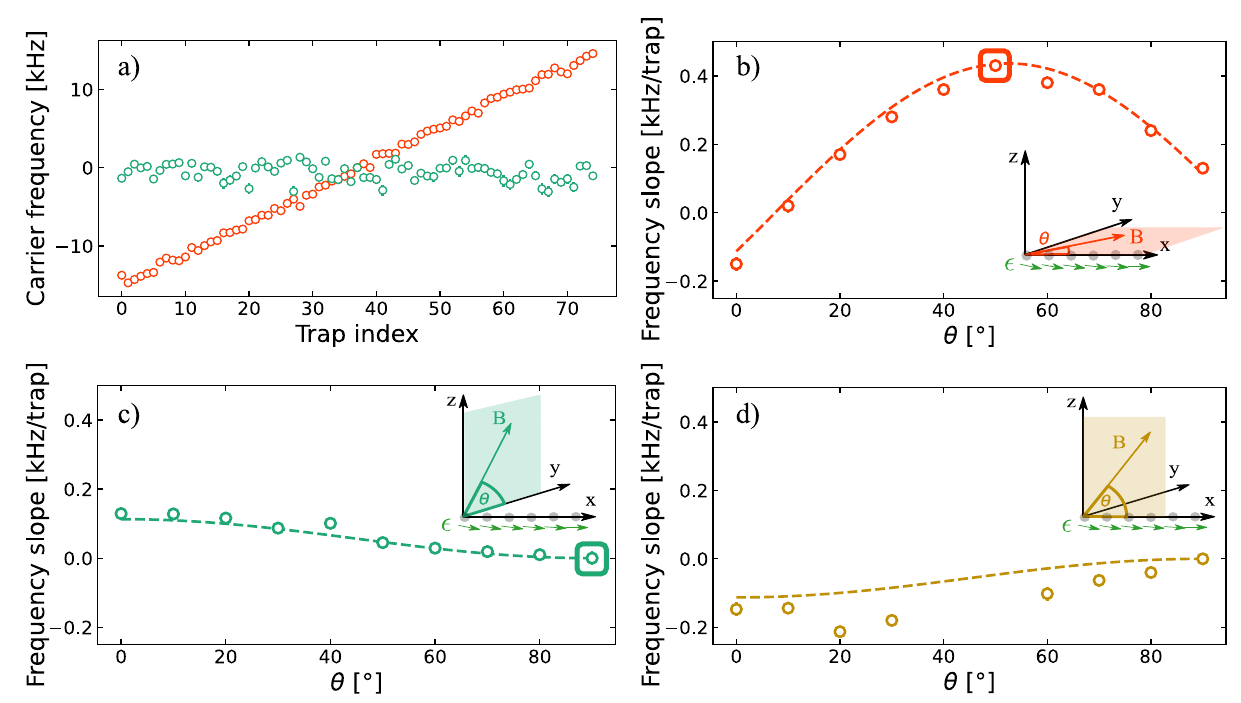}
    \caption{Spatial frequency gradients in the Raman transition. a) Trap-resolved carrier frequency extracted from Raman spectra with the $\bm{B}$ field in the $xy$ plane and $\theta = 50^\circ$ (red) or along $z$ (green). b-c-d) Frequency slope in the Raman spectra for different orientations of the $\bm{B}$ field. The angle $\theta$ is defined with respect to $x$ ($xy$ and $xz$ planes) or with respect to $y$ ($yz$ plane), corresponding to \textbf{B} = $(B\cos\theta, B\sin\theta, 0)$ (panel b), \textbf{B} = $(0, B\cos\theta, B\sin\theta)$ (panel c) and \textbf{B} = $(B\cos\theta, 0, B\sin\theta)$ (panel d). The $x$ axis contains the atom array and $y$ is the propagation direction of the tweezers beam. Dashed lines are calculations assuming a gradient of the tweezers polarization direction in the $xy$ plane (see text), schematically shown with the green arrows. The boxed points correspond to the spectra in panel a).}
    \label{fig:RamanSlopes}
\end{figure*}

Due to the tensorial light shift, we would expect fluctuations in the Raman resonance along the array, similar to the measurements in Fig. \ref{fig:4} for the $\ket{g} \rightarrow \ket{e}$ transition. Interestingly, instead, we measure a constant energy gradient along the array that depends on the orientation of the magnetic field. The measurements are summarized in Fig. \ref{fig:RamanSlopes}. We fix the magnetic field magnitude $B = 3.3$ G and we change its direction in the three orthogonal planes $xy$, $xz$ and $yz$ . The $x$ axis contains the atomic array and the $y$ axis is the propagation direction of the tweezers beam. Depending on the value of the magnetic field angle $\theta$, we observe energy gradients that can be both positive or negative, going up to a total energy shift over the 75 traps equal to 30 kHz (for $P_T = 1.8$ mW, shown in Fig. \ref{fig:RamanSlopes}(a)). Note that here, contrary to the main text, we rotate $\textbf{B}$ also in the $xy$ plane. In this case, $\theta$ is the angle between $\textbf{B}$ and the $x$ axis. We identify two configurations in which the slope vanishes: in the $xy$ plane around $10^\circ$ and along the $z$ axis. These configurations are favorable because they allow averaging over the 75 traps, speeding up the data acquisition time. The data plotted in Fig. \ref{fig:3}(b) are taken with the magnetic field along the $z$ axis. \\

We checked that the energy variation is indeed due to a varying light shift, by verifying that it is linear with the trap power (we therefore exclude magnetic field gradients). The observed light shift gradients can be indeed explained by a polarization gradient of the trapping light in the array, which generates a position-dependent shift of the Raman resonance through the tensor polarizability. Trap-resolved energy shifts due to a trap-dependent polarization have already been reported with strontium in a square 2D tweezers grid \cite{Am24}. The full expression of the light shift hamiltonian is 
\begin{align}\label{eq:LightShift}
\begin{split}
\hat{H}_{LS} = &
-\frac{I}{2\epsilon_0 c}\Big[\alpha_s - i\alpha_v(\bm{\epsilon}^* \times \bm{\epsilon}) \hat{\bm{J}}/2J +\\
& + \alpha_t \Big(\frac{\{\bm{\epsilon}\cdot\bm{\hat{J}} , \bm{\epsilon}^*\cdot\bm{\hat{J}}\}-2\bm{\hat{J}}^2}{2J(2J-1)} \Big)   \Big],
\end{split}
\end{align}
where $\alpha_s$, $\alpha_v$ and $\alpha_t$ are the scalar, vector and tensor polarizabilities, $\bm{\epsilon}$ the polarization vector, and $\{\hat{A}, \hat{B} \} = \hat{A}\hat{B} + \hat{A}\hat{B}$. With a large magnetic field, the light shift hamiltonian reduces to an energy shift of the Zeeman eigenstates, and $m_J$ is still a good quantum number. The state-dependent light shift comes from the vector (linear in $m_J$) and tensor (quadratic in $m_J$) parts.\\

As discussed is Section \ref{sec:3}, even if the vector contribution is non zero due to the small ellipticity angle of the tweezers polarization, we expect it to be much smaller than the tensor contribution. We can then qualitatively understand the measurements in Fig.~\ref{fig:RamanSlopes} considering only the tensor contribution and zero ellipticity, as in Eq.~\eqref{eq:polarizability} in the main text. To do so, we consider traps with a linear polarization in the $xy$ plane that depends on the trap position $x$: $\epsilon (x) = (\cos(\gamma_0+\gamma'\,x),\sin(\gamma_0+\gamma'\,x),0)$. In this case, when the field is also in the $xy$ plane with angle $\theta$ as in Fig.~\ref{fig:RamanSlopes}(b), the light shift depends on $(\bm{\epsilon} \cdot \bm{b})^2=\cos^2(\theta-(\gamma_0+\gamma'\,x))$. Its variation with $x$, at first order in $\gamma '$, is proportional to $\gamma'\,\sin(2(\theta-\gamma_0))$, with a maximum at $\theta-\gamma_0=\pi/4$. This reproduces well the measurements in Fig.~\ref{fig:RamanSlopes}(b), with $\gamma_0\approx7^\circ$, indicating that the magnetic field is not perfectly aligned with the average polarization vector when oriented along $x$.
When instead $\bm{b}$ is along the $z$ axis, $(\bm{\epsilon} \cdot \bm{b})^2 = 0$ and we expect no gradient of the light shift in agreement with the results shown in Fig.~\ref{fig:RamanSlopes}.\\

The simple modelization above explains qualitatively the results. To quantitatively compare with the measurements for all angles $\theta$, in Fig. \ref{fig:RamanSlopes}(b-c-d), we compute the full expression of the light shift in Eq. \ref{eq:LightShift} including also the small ellipticity and vector contribution. We find the best fit with the data with a tensor polarizability $\alpha_t = -20$ au (in agreement with \cite{Bl24}) a mean polarization centered on $\gamma_0=7.5^\circ$ and a gradient $\gamma ' = \SI{1.4}{mrad\per\micro\meter}$, corresponding to $15^\circ$ over the 75 traps. The results are plotted as dashed lines in Fig.\,\ref{fig:RamanSlopes}(b-c-d). We find good agreement with the data. \\
The gradient of polarization is likely due to a propagation direction which differs from trap to trap. We have not yet investigated how this gradient can be compensated experimentally. 

\bibliography{biblio}

\end{document}